\documentstyle[12pt]{article}
\begin{document}
\newcommand {\be}{\begin{equation}}
 \newcommand {\ee}{\end{equation}}
 \newcommand {\bea}{\begin{array}}
 \newcommand {\ben}{\begin{eqnarray}}
 \newcommand {\een}{\end{eqnarray}}
 \newcommand {\cl}{\centerline}
 \newcommand {\eea}{\end{array}}
 \renewcommand {\theequation}{\thesection.\arabic{equation}}
 \newcommand {\newsection}{\setcounter{equation}{0}\section}
  \baselineskip 0.6 cm
\begin{titlepage}
\vspace{-10mm}

\begin{center}
\begin{large}
{\bf Perfect and Imperfect  \\  Gauge Fixing }\\
\end{large}
\vspace{5mm}
 {\bf A. Shirzad \footnote{e-mail:
shirzad@ipm.ir}}

\vspace{12pt}
{\it Institute for Studies in Theoretical Physics and Mathematics\\
P. O. Box: 5531, Tehran, 19395, IRAN, \\Department of  Physics,
Isfahan University of Technology \\
Isfahan,  IRAN. } \vspace{0.3cm}
\end{center}
\abstract{Gauge fixing may be done in different ways. We show that
using the chain structure to describe a constrained system,
enables us to use either a perfect gauge, in which all gauged
degrees of freedom are determined; or an imperfect gauge, in which
some first class constraints remain as subsidiary conditions to be
imposed on the solutions of the equations of motion. We also show
that the number of constants of motion depends on the level in a
constraint chain in which the gauge fixing condition is imposed.
The relativistic point particle, electromagnetism and the Polyakov
string are discussed as examples and perfect or imperfect gauges
are distinguished.

 \vfill}
\end{titlepage}

\section{Introduction}
There are two well-known methods to construct the constraint
structure of a constrained system \cite{dirac,HENOUX}. First, the
{\it level by level} method \cite{BGPR} in which the equations
concerning the consistency of constraints at a given level are
solved simultaneously to find the constraints of the next level.
Second, the {\it chain by chain} method \cite{lorshir} in which
the consistency of every primary constraint produces the
corresponding constraint chain up to the end. In the second method
the constraints are organized in separate first and second class
chains. As is well-known, the first class constraints (FCC's) are
generators of gauge transformations which correspond to the
emergence of arbitrary functions of time in the solutions of
equations of motion. The relationship between the first class
constraints, the generating function of gauge transformations and
the arbitrary functions of time has intensively been studied in
the literature
\cite{shirshab,HenTeiZan,GracPons,GomHenPon,ChaiMart}. However,
this relationship can be better understood in the context of the
chain by chain method. Every first class constraint chain of $N$
entries corresponds, in the solutions of equations of motion, to
an arbitrary function of time together with its derivatives up to
the $(N-1)$th level.

In the presence of gauge transformations any physical state
corresponds to an orbit in the phase space, i.e. the gauge orbit,
along which only the arbitrary functions of time do change. Gauge
transformations, generated by FCC's, just translate the system
along the gauge orbits, without changing the physical state. Gauge
fixing means that coordinates describing the gauge orbits are
determined so that a one to one correspondence holds between the
physical states and the points of the remaining subspace of the
phase space. In this way it is needed to impose, by hand, extra
constraints on the system to fix the gauges. We call these
constraints as ``the gauge fixing conditions" (GFC's).

Suppose for simplicity that we have chosen some suitable
coordinates in which the FCC's are converted to some momenta. Then
gauge fixing is equivalent to determining the coordinates
conjugate to the FCC's. This means that the GFC's should have
non-vanishing Poisson brackets, at least, with a subset of FCC's.
When the gauges are fixed, there exist no more arbitrary functions
of time (or arbitrary fields in the case of a field theory) in the
solutions of the equations of motion.

However, for the sake of consistency, and depending on the way
GFC's are chosen, some FCC's may still remain as additional
necessary conditions which should be imposed on the physical
solutions of the problem. This feature, though encountered  for
instance in string theory, has not been recognized so far in the
context of constrained systems. For example, one effect of this
method concerns the number of initial constants appearing in the
solutions of the equations of motion, which will be discussed in
this paper.

In the following section we will first review the basic concepts
of the chain structure and the proposal of Ref. \cite{shirlor}
which implies {\it perfect gauge fixing} of a given gauge theory.
In this article we want to show that this is only one possibility.
In fact, it is also possible to fix the gauges in an {\it
imperfect} way. Analyzing a simple toy model in section (2) we
will explain the main idea of the paper that the details of gauge
fixing (including the number of initial conditions) depend on the
definite level in a constraint chain in which the GFC is imposed.

Section (3) is devoted to investigating the relativistic point
particle. The constraint structure, the gauge transformation
(which in this case is reparametrization) and the corresponding
generating function, perfect and imperfect gauge fixing and
finally the number of initial constants in this problem are
discussed. The next interesting model is electromagnetism, which
is studied in section (4) mostly in relationship with the problem
of gauge fixing. We will show that the famous Coulomb gauge is a
perfect gauge, while the Lorentz gauge is of a completely
different nature which we call a {\it completely imperfect gauge}.

In section (5) we will investigate the constraint structure of the
bosonic string theory and analyze different gauges traditionally
used in the literature. We show that the famous covariant gauge is
an imperfect one and implies imposing the Virosoro constraints as
subsidiary conditions on the solutions of the equations of motion
as well as on the physical states in the quantum theory. On the
other hand, the light cone gauge, although disturbing the manifest
covariance of the theory, is a perfect gauge which preserves only
purely physical degrees of freedom. In section (6) we will give
our concluding remarks.

\section{Gauge fixing in chain structure}

Suppose, for simplicity, that we have just one primary constraint,
$\phi_{1}$, in a system described by the canonical Hamiltonian
$H_c$. The dynamics of every function $g(q,p)$ is achieved by
 \be \dot{g}(q,p)=\left\{ g,H_t \right\} \label{a1} \ee
where $H_t$ is the total Hamiltonian given as
 \be H_t=H_c+\lambda \phi_{1} \label{a2}, \ee
 in which $\lambda$ is the undetermined Lagrange multiplier.
Following the conventional consistency procedure of Dirac
\cite{dirac}, i.e. $\dot{\phi}\approx 0$ (where $\approx$ means
weak equality), the second level constraint emerges as
 \be \phi_{2}=\left\{ \phi_{1},H_c \right\}\approx 0.\label{a3} \ee
We are interested in the first class systems where $\left\{
\phi_{2},\phi_{1} \right\}\approx 0$. Therefore, the consistency
of $\phi_2$, from Eq. (\ref{a1}), gives $\phi_3=\left\{
\phi_{2},H_c \right\}$ and so on. In this way a constraint chain
is derived via {\it the chain rule}
 \be \phi_{n+1}=\left\{ \phi_{n},H_c \right\}, \label{a4}  \ee
provided that $\left\{ \phi_n, \phi_1 \right\} $ vanish (at least
weakly) so that the Lagrange multiplier $\lambda$ is not
determined at any stage. A first class chain terminates at level
$N$, say, where
 \be \left\{ \phi_{N},H_c \right\}\approx 0. \label{a5} \ee
Hence, the corresponding Lagrange multiplier remains undetermined.
It can be shown that the solutions of the equations of motion in
this case contain one arbitrary function of time and its time
derivatives up to order $N-1$ \cite{HENOUX}. For instance, in the
simple case where $\left\{ \phi_{N},H_c \right\}$ as well as
$\{\phi_n,\phi_1\}$ for all $n$ vanish strongly, it is shown
\cite{shirshab} that the gauge transformations are generated by
the following function
 \be G=\sum_{s=1}^{N} (-1)^{s}\, \phi_{s}(q,p) \,
  \frac{d^{N-s}\eta (t) }{dt^{N-s}},  \label{a6} \ee
where $\eta (t)$ is an infinitesimal arbitrary function of time.

The above considerations can be easily generalized to a
multi-chain system where one should only add a chain label to
constraints as well as arbitrary functions of time. The number of
chains, arbitrary functions of time, and the primary first class
constraints are the same. However, the chains may have different
lengths. It should be noticed that for a generic system it is not
an easy task to arrange the constraints as chains. In fact, this
requires a special algorithm to be followed as given in Ref.
\cite{lorshir}.

Now let us see how the gauge can be fixed. Since all the first
class constraints $\phi_{n}$ are generators of gauge
transformation, it may seem that one should impose the same number
of GFC's as that of constraints, i.e. the GFC $\omega_{n}=0$
should be imposed to fix the gauge transformation generated by
$\phi_{n}$. However, the key point is that the GFC's should remain
valid in the course of time in the same way as the FCC's
themselves. This fact brings our attention to two points: first,
if the GFC's are not chosen appropriately, their consistency may
lead to extra constraints which may {\it overdetermine} the
system; second, one may shorten the way through finding all GFC's
needed to fix the gauge by giving a smaller number of GFC's and
finding the rest of them by following their consistency
conditions. This is in fact the main idea of Ref. \cite{shirlor},
where the authors proposed imposing the {\it primary} GFC
$\omega_{N} \approx 0$ where $\omega_{N}$ is conjugate to the
terminating element of the chin (while commuting with the others),
i.e.
 \be \left\{ \omega_{N} ,\phi_{n} \right\} =
 \chi (q,p)\delta_{n,N}, \label{a7}  \ee
where $\chi$ should not vanish on the surface of the constraints.

To get a better idea of how this method works suppose that the
terminating element is one of the momenta, say $p_k$. Clearly the
conjugate coordinate $q_k$ is not contained in the previous
constraints (otherwise we would not have a first class system).
Then from (\ref{a6}) the gauge transformation of $q_k$ is just
$\delta q_k=\eta(t)$. In other words $q_k$ is an arbitrary
function depending on the gauge. Once this function is chosen by
the gauge $\omega_N = q_k - f(t) \approx 0$, where $f(t)$ is some
given function of time, the gauge would be fixed completely. Since
$\omega_N$ is an explicit function of the time its consistency
leads to the next GFC via the formula
 \be \dot{\omega}_{N}\equiv \omega_{N-1}= \left\{\omega_{N},H_c
 \right\}+\frac{\partial\omega_{N}}{\partial t}.\label{aa1}\ee
Using the chain rule (\ref{a4}) and the Jacobi identity one can
show that $\omega_{N-1}$ is conjugate to $\phi_{N-1}$, and so on.
Hence, the GFC's in turn obey the chain rule
 \be  \omega_{n-1}=\left\{ \omega_n , H_c \right\}
 +\frac{\partial^{N-n+1}f}{\partial t^{N-n+1}} \label{a8}  \ee
and constitute conjugate pairs with FCC's:
 \be \left\{ \omega_n,\phi_n \right\}=(-1)^{N-n}\chi . \label{a9} \ee
The procedure goes on up to the last step where consistency of
$\omega_1$ determines the Lagrange multiplier as
 \be \lambda= \frac{ (-1)^N\left\{
 \omega_1,H_c \right\}+\partial^{N}f/\partial t^{N+1}}
 {\chi(q,p)}\ .  \label{a10}   \ee
The above procedure, which we call {\it perfect gauge fixing},
leads to a complete fixing of the gauge. The reduced phase space
achieved by imposing the whole FCC's and GFC's has the dimension
of $2K-2N$ where $2K$ is the dimension of the original phase
space. Therefore, the number of physical degrees of freedom (which
come through second order differential equations of motion) would
be $K-N$. In this way, in perfect gauge fixing, the number of
constants to be determined by the initial conditions is $2(K-N)$.
For a multi-chain system this would be clearly $2(K-\sum_a N_a)$,
where $a$ is the chain index.

Now let us see what happens if the gauge fixing does not begin
from the terminating element of the chain. We call such a method
as {\it imperfect gauge fixing}. Suppose, for some reason, one has
begun fixing the gauge from some intermediate element in the
chain, say from $\phi_M$, where $M < N$. By this we mean that one
imposes the GFC $\omega_M$, instead of $\omega_N$, such that
 \be \left\{ \omega_{M} ,\phi_{n} \right\} =
 \chi (q,p)\delta_{n,M}. \label{a11}  \ee
Note specially that $\omega_M$ commutes with the constraints
succeeding $\phi_M$ as well as the ones preceding it. Then the
consistency process gives the set of GFC's $\omega_{M-1},
\omega_{M-2} \cdots $, similar to perfect gauge fixing. At the
last step $\lambda$ is determined similar to Eq. (\ref{a10}) with
$N$ replaced by $M$. In this way the set $\phi_1,\cdots,\phi_M,
\omega_M, \cdots,\omega_1$ serves as a system of second class
constraints which leads to a reduced phase space with dimension
$2K-2M$.

However, we are leaved with the constraints
$\phi_{M+1},\cdots,\phi_N$, which are not yet fixed during the
gauge fixing process. Although the gauge is fixed so that there
remains no arbitrary function of time in the solutions of
equations of motion, {\it one should still impose the remaining
constraints $\phi_{M+1},\cdots,\phi_N$ on the solutions} to get a
consistent physical system. In other words, the classical
solutions are achieved by solving second order differential
equations for $K-M$ variables together with imposing $N-M$
constraints (appearing in the shape of first or zeroth order
differential equations in configuration space). Therefore, the
number of constants to be determined by initial conditions is:
$2K-2M-(N-M)=2K-N-M$.

Imperfect gauge fixing has also  considerable effects on
quantization procedure. We remind that there are two methods for
quantizing a first class system. The first one is to fix the
gauges completely and then quantize the  reduced phase space
variables by converting them to operators and their Dirac brackets
to commutators. The second method is to quantize all the original
phase space variables by converting the original Poisson brackets
to commutators and then impose the condition
 \be \textrm{FCC} |\textrm{phys}\rangle=0,  \label{a12} \ee
where $|\textrm{phys}\rangle$ means ``physical states". The reason
for this condition is the generator of gauge transformations in
the general case can be written in terms of first class
constraints. Hence, Eq. (\ref{a12}) results from the physical
condition $G|\textrm{phys}\rangle=0$.

The quantization procedure in an imperfect gauge fixed system is a
mixture of both methods. In this case, the variables of the
$2K-2M$ dimensional reduced phase space should first be quantized
by converting the following Dirac brackets to commutators,
 \be \{f,g\}_{\scriptstyle{DB}}=\{f,g\}-\left\{f,\psi_r\right\}
 C^{rs}\left\{\psi_{s},g\right\},  \label{a14} \ee
where
 \be \psi_{r} ,\psi_{s} \;\epsilon\left\{ \phi_{1}, \cdots
 \phi_M,\omega_M, \cdots,\omega_1 \right\} \label{a15} \ee
and $C^{rs}$ is the inverse of
 \be C_{rs}= \left\{ \psi_{r},\psi_{s}\right\}. \label{a16} \ee
 Then the following condition should be imposed on states to
achieve the physical ones,
 \be \widehat{\phi_n}\,|\textrm{phys}\rangle=0
 \hspace{1cm} M+1<n<N \label{a13} \ee
where $\widehat{\phi_n}$ is the operator version of the constraint
$\phi_n$.

To see the above ideas more clearly consider a simple toy model
with $(x,y,z)$ as the variables, described by the Lagrangian
 \be  L=\dot{x}\dot{y}-y z.  \label{b1}  \ee
The momentum $p_z$ emerge as the primary constraint. The total and
canonical Hamiltonian read
 \be \bea{l}H_t=H_c+\lambda p_z,\\H_c=p_xp_y+yz.\eea\label{b2} \ee
Using the chain rule (\ref{a4}), the following first class
constraint chain is derived
 \be \bea{l} \phi_1=p_z \\ \phi_2=-y \\ \phi_3=-p_x. \eea
 \label{b3} \ee
Since the last element of the chain commutes strongly with $H_c$,
the generator of gauge transformation can be written from Eq.
(\ref{a6}) as
 \be G=-p_z\ddot{\eta}-y\dot{\eta}+p_x\eta. \label{b4} \ee
Suppose we want to fix the gauge perfectly. This is done by
imposing the GFC
 \be \omega_{3}=x-f(t). \label{b5} \ee
Using Eq. (\ref{a8}), the consistency of $\omega_3$ gives the next
two GFC's as
 \be \bea{l} \omega_2=p_y-\dot{f}(t) \\ \omega_1=-z-\ddot{f}(t)
 \eea \label{b6} \ee
Consistency of $\omega_1$, using the total Hamiltonian, determines
the Lagrange multiplier as
 \be \lambda=-\frac{d^3f}{dt^3}\; . \label{b7} \ee
As can be seen, this system with three degrees of freedom, obeys
three first class constraints, which means that the system is
completely gauged (has no further dynamical degree of freedom).
So, by a perfect gauge fixing there remains no dynamics in the
system. In other words, all the variables are determined by the
choice of the function $f(t)$. Moreover, since $N=K=3$, the number
of initial constants is zero.

Now let us do an imperfect gauge fixing in this system. Suppose
one prefers to fix the gauge by imposing the GFC
 \be \omega'_1 =z-g(t), \label{b8} \ee
which fixes the value of $z$ whose variation is generated by the
FCC $p_z$. Consistency of  $\omega'_1$, using Eq. (\ref{aa1}),
determines the Lagrange multiplier  as
 \be \lambda=\dot{g}(t). \label{b9} \ee
The total Hamiltonian turns out to be
 \be H_t=p_xp_y+g(t)y+\dot{g}(t)p_z. \label{b10} \ee
The four dimensional reduced phase space acquires the following
equations of motion
 \be \bea{ll} \dot{x}=p_y, \hspace{1.5cm} &\dot{p}_y=-g(t) \\
 y=0 &p_x=0 \eea \label{b11} \ee
Equations in the first line are derived from the total Hamiltonian
(\ref{b10}), while the ones in the second line are the constraints
remained at the tail of the constraint chain (\ref{b3}) without
imposing corresponding GFC's. In this example these two sets of
equations are separated; this point is not essential for the
general case. The final solution of the equations of motion for
the remaining two degrees of freedom, i.e. $x$ and $y$, are
obtained as
 \be \ddot{x}=-g(t), \hspace{1.5cm} y=0. \label{b12} \ee
Integrating the first equation brings in two constants, in
agreement with the formula $2K-N-M=6-3-1=2$.

The above considerations can be seen more clearly in the
Lagrangian framework. The Euler-Lagrange equations of motion due
to the Lagrangian (\ref{b1}) read
 \be \bea{l} \displaystyle{\frac{\delta L}{\delta x}}=\ddot{y}=0
 \vspace{2mm}\\ \displaystyle{\frac{\delta L}{\delta y}}=
 \ddot{x}+z=0 \vspace{2mm} \\ \displaystyle{\frac{\delta L}
 {\delta z}}=y=0 \eea \label{b13} \ee
The first equation can result from the third one, which requires
that $y$ is fixed at zero. The remaining equation constrains the
time behavior of $x$ and $z$. If one determines $x$ as the given
function $f(t)$, then $z$ would be completely determined as
$\ddot{f}(t)$. Conversely, if one determines $z$ as a definite
function $g(t)$, then $x$ should be found by integrating $g(t)$
twice which brings in two constants of integration.

Finally let us take a look at the problem of quantization of the
model. In perfect gauge fixing the reduced phase space is null and
no degree of freedom is remained to be quantized. On the other
hand, in imperfect gauge fixing the canonical operators
$(\hat{x},\hat{p}_x,\hat{y},\hat{p}_y)$ describe a quantum
particle in two dimensions. However, the physical subspace due to
the conditions (\ref{a13}) is restricted to the states satisfying
 \be \hat{y}|\psi>=\hat{p}_x|\psi>=0. \label{b14} \ee
These two conditions are so powerful to kill all the states except
the single state $|\psi>=|y=0,p_x=0>$ with the wave function
 \be \psi(x,y)=\frac{1}{2\pi}  \delta(y).
 \label{b15} \ee

\section{Relativistic Point Particle}

Consider a relativistic point particle in a $D$-dimensional
Minkovski space-time described by the action
 \be  S=\frac{1}{2}\int d\tau \left( \eta^{-1}
 \dot{X}^{\mu}\dot{X}_{\mu}-m^{2}\eta \right) \label{c1} \ee
where $m$ is the mass of the particle, "dot" means differentiating
with respect to $\tau$, the proper time, and $\eta(\tau)$ is an
auxiliary variable called the {\it ein-bin} variable. The
canonical momenta conjugate to $X^{\mu}$ and $\eta$ are
respectively
 \be P_{\mu}=\eta^{-1}\dot{X}_{\mu}, \hspace{1cm} P_{\eta}=0.
 \label{c2} \ee
So $P_{\eta}$ is the primary constraint. The canonical and total
Hamiltonian are as follows
 \be \bea{l} H_c=\frac{1}{2} \left( \eta
 P^{\mu}P_{\mu}+m^{2}\eta \right), \\
 H_t=H_c+\lambda P_{\eta}.  \eea \label{c3}\ee
The consistency process gives the following constraint chain
 \be \bea{l} \phi_{1} = P_{\eta}\\ \phi_{2} = -
 \frac{1}{2} \left( P^{\mu}P_{\mu}+m^2
  \right)\; . \label{c5} \eea \ee
Since $\left\{ \phi_{2}, H_t \right\}$ vanishes strongly, the
generator of gauge transformation, using (\ref{a6}), can be
written in terms of an arbitrary infinitesimal function
$\epsilon(t)$ as
 \be G=-\dot{\epsilon}\phi_{1}+\epsilon \phi_{2} \label{c7} \ee

Let us see which transformation $G$ generates. Using
(\ref{c5}-\ref{c7}), the variations of $X^{\mu}$ and $\eta$ under
the action of $G$ are respectively
 \ben \delta X^{\mu} &\equiv& \left\{ X^{\mu},G\right\}=
 \epsilon P^{\mu} \label{c8}\\ \delta\eta &\equiv& \left\{
 \eta, G\right\}=\dot{\epsilon}. \label{c9} \een
Using the definition of $P_{\mu}$, Eq. (\ref{c8}) gives
 \be \delta X^{\mu}=\epsilon\,\eta^{-1}\dot{X}^{\mu} \label{c10}\ee
Eq. (\ref{c9}) shows that $\eta(\tau)$ is somehow arbitrary.
Therefore, assuming $\xi(\tau)\equiv-\epsilon\eta^{-1}$; we have
 \ben \delta X^{\mu}&=&-\frac{dX^{\mu}}{d\tau}\,\xi(\tau),
 \label{c11}\\ \delta\eta &=& -\dot{\eta}\,\xi-\eta\,\dot{\xi}\, .
 \label{c12} \een
It is easily seen that the action (\ref{c1}) is invariant under
the reparametrization
 \be \tau \rightarrow \tau'=\tau+\xi(\tau), \label{c13} \ee
provided that the transformed variables behave as follows
 \ben X'^{\mu}(\tau') &=&X^{\mu}(\tau), \label{c14}\\
 \eta'(\tau')d\tau' &=& \eta(\tau)d\tau. \label{c15} \een
Now we show that the variations derived in Eqs. (\ref{c11}) and
(\ref{c12}) correspond to an infinitesimal reparametrization. To
do this, using Eq. (\ref{c14}) we can write
 \ben \delta X^{\mu} &\equiv & X'^{\mu}(\tau)-X^{\mu}(\tau)
 \nonumber\\ &\cong&X'^{\mu}(\tau)-X'^{\mu}(\tau')\nonumber \\
  &\cong& -\frac{\partial X'^{\mu}}{\partial \tau}\delta\tau\nonumber\\
  &\cong&-\frac{\partial X^{\mu}}{\partial \tau}\xi(\tau),\nonumber
  \een
where $\cong$ means equality up to the first order quantities in
terms of the infinitesimal variables. On the other hand, Eqs.
(\ref{c15}) and (\ref{c13}) imply that
 $$ \eta'(\tau')(1+\dot{\xi})=\eta(\tau), $$
which gives
 \ben \delta\eta &\equiv& \eta'(\tau)-\eta(\tau)\nonumber\\
 &\cong& \eta'(\tau')-\eta(\tau') \nonumber\\ &\cong&
 \eta(\tau)-\dot{\xi}(\tau)\eta(\tau)-\eta(\tau')\nonumber\\
 &\cong& -\frac{\partial\eta}{\partial\tau}d\tau-
 \dot{\xi}(\tau)\eta(\tau)\nonumber \\ &\cong& -\dot{\eta}(\tau)
 \xi(\tau)-\dot{\xi}(\tau)\eta(\tau). \nonumber  \een
These calculations show that the gauge generating function $G$ in
Eq. (\ref{c7}) is in fact the generator of infinitesimal
reparametrizations.

Now let us proceed to the problem of gauge fixing. It is clear
that at most one arbitrary function of time would appear in the
solutions of the equations of motion. Therefore, different gauges
correspond to the choice of the variable which is determined by
the gauge (e.g. one of the $X^{\mu}$'s or $\eta$). One simple
choice is considering $\eta$ as the given function $f(t)$ by
imposing the GFC $\omega'_1=\eta-f(t)$. Since this gauge fixes
only the first entry in the constraint chain (\ref{c5}), it is an
imperfect gauge. The consistency of $\omega'_1$ from (\ref{aa1})
determines the Lagrange multiplier $\lambda$ in (\ref{c3}) as
$\lambda=\dot{f}$. The canonical equations of motion read
 \be \bea{l}\dot{X}^{\mu} = \eta P^{\mu} \\
 \dot{P^{\mu}} = 0. \label{c18} \eea \ee
Eqs. (\ref{c18}) together with the remaining constraint
$\phi_2=P_{\mu}P^{\mu}+m^2=0$, determine all the variables. It is
easy to see that these equations bring in $2D-1$ constants of
integration, in agreement with the formula $2K-N-M=2(D+1) -2-1=
2D-1$.

It is worth to note the Lagrangian equations of motion, i.e.
 \ben \frac{\delta L}{\delta \eta} &=& -\eta^{-2} \dot{X}^{\mu}
 \dot{X}_{\mu}-m^2=0, \label{c21} \\ \frac{\delta L}{\delta
 X^{\mu}} &=& \frac{d}{d\tau}\left( \eta^{-1}\dot{X}_{\mu}
 \right)=0. \label{c22} \een
Eq. (\ref{c21}) is acceleration-free and serves as a first level
(as well as last level) Lagrangian constraint. \footnote{We remind
that the $n$th level Lagrangian constraint corresponds to
$(n-1)$th level Hamiltonian one \cite{BGPR}.} It is easily seen
that the Lagrangian equations of motion (\ref{c21}) and
(\ref{c22}) are not sufficient to determine all the variables.
However, assuming $\eta(t)$ as an arbitrary function, we can
determine $X^{\mu}$'s in terms of $\eta(t)$ and the constants
$P^{\mu}$ by integrating the equations $P^{\mu}=\eta^{-1}
\dot{X}^{\mu}$. The number of independent $P^{\mu}$'s is $D-1$
according to the condition $P^{\mu}P_{\mu}+m^2=0$ resulting from
(\ref{c21}). Integrating $\dot{X}^{\mu}=\eta(t)P^{\mu}$ brings in
$D$ further integration constants, adding up to $2D-1$ as
expected.

One can consider a perfect gauge by imposing a desired time
dependence for one of the $X^{\mu}$'s or a combination of them.
The most famous gauge is the temporal one, in which $X^0$ is
assumed to be the same as the proper time. The primary GFC in this
gauge is
 \be \omega_2=X^0-\tau. \label{c23} \ee
Using (\ref{aa1}) the consistency of $\omega_2$ gives the next GFC
as
 \be \omega_1= \eta P^0-1. \label{c24} \ee
The set of canonical equations (\ref{c18})  together with the
constraints $\phi_1$ and $\phi_2$ and the GFC's $\omega_2$ and
$\omega_1$ determine all the variables as
 \be \bea{lll} \eta &=& \displaystyle{\frac{1}{P^0} }\\ X^0 &=& \tau
 \\ X^i &=& \displaystyle{\frac{P^i}{P^0}\,\tau+x^{0i}}
  \label{25}\eea \ee

The total number of constants in this case is $2(D-1)$ where $D-1$
of them are the independent $P^{\mu}$'s (remember the constraint
$\phi_2$ implies ${P^0}^2=\sum {P^i}^2+m^2$) and $D-1$ of them are
the $x^{0i}$'s. This is in agreement with the formula $2K-2N=
2(D+1)-4=2D-2$.

Similar treatment can be done in the light cone coordinates where
$X^{\pm}\equiv(X^0\pm X^1)/\sqrt{2}$ and $X^i\equiv X^\mu \ \ \
\mu =2,\cdots D$. A perfect gauge fixing in these coordinates can
be achieved by imposing the GFC $\omega_2=X^+-\tau=0$ whose
consistency gives $\omega_{1}=-\eta P_--1=0$. The reduced phase
space is achieved by omitting the canonical pairs
$(\eta,P_{\eta})$ and $(X^+,P_+)$. Independent variables $X^i$ and
$X^-$ can be solved in terms of $2(D-1)$ constants
$P_i,P_-,X^{0i}$ and $X^{0-}$ as
 \be \bea{l} \displaystyle{X^i(\tau)=-\frac{P_i}{P_-}\tau+X^{0i}}
 \vspace{2mm}\\ \displaystyle{X^-(\tau)=-\frac{\sum
 P_i^2+m^2}{P_-^2}}\tau+X^{0-}\ . \label{c26} \eea \ee

\section{Electromagnetism}

Consider the famous action of the electromagnetism as
 \be S=-\frac{1}{4}\int d^{4}xF_{\mu\nu}F^{\mu\nu}, \label{d1} \ee
where $F^{\mu\nu}\equiv \partial^{\mu} A^{\nu}- \partial^{\nu}
A^{\mu}$. The canonical momenta are $\Pi_{\mu}=-F_{0\mu}$ which
yield $\phi_{1}=\Pi_{0}$ as the primary constraint. The total
Hamiltonian reads
 \be H_t=H_c+\int d^{3}x\lambda(x)\Pi_{0}(x), \label{d2} \ee
where $H_c$ is the canonical Hamiltonian;
 \be H_c=\int d^{3}x\left[\frac{1}{2}\Pi_{i}\Pi_{i}+
 \frac{1}{4}F_{ij}F_{ij}-A_{0}\partial_{i}\Pi_{i} \right].
 \label{d3} \ee
We assume the metric to be $g_{\mu\nu}= \textrm{diag} (-+++)$.
Consistency of $\phi_{1}$ gives the secondary constraint
$\phi_{2}=\partial_{i}\Pi_{i}$. Consistency of $\phi_2$ is
fulfilled identically. So we have a constraint chain with two
elements.

A perfect gauge fixing can be achieved by imposing $\omega_{2} =
\partial_{i}A^{i}$ as the primary GFC which is conjugate to
$\phi_{2}$. Consistency of $\omega_{2}$ gives the next GFC as
$\omega_{1}=\partial_{i}\Pi_{i}+\partial_{i}\partial_{i}A_{0}$
which is weakly equivalent to $\partial_{i}\partial_{i}A_{0}$.
Finally consistency of $\omega_{1}$ determines $\lambda$ as any
function with vanishing divergence. A well defined Dirac bracket
would emerge from the second class set given by
$\phi_{1},\phi_{2},\omega_{2}$ and $\omega_{1}$ which is
well-known in the literature \cite{sunder}. This gauge is the
famous Coulomb gauge.

One can also perform an imperfect gauge by imposing the GFC
$\omega'_{1}=A^{0}$ which is conjugate to $\phi_1$. This gauge
determines the Lagrange multiplier $\lambda(x)$ to be identical to
zero which yields $H_t=H_c$. Even though this choice of gauge
kills the arbitrariness of the theory, we are still remained with
the not yet fixed constraint $\phi_2$. The canonical equations of
motion read
 \ben \dot{A}_{i}&=&\Pi_{i}+\partial_{i}A_{0} \label{d4} \\
 \dot{\Pi}_{i} &=& \partial_{j}F_{ji} \label{d5} \een
Eliminating the canonical pair $(A^{0},\Pi_{0})$, determines the
rest of the variables via the Eqs. (\ref{d4}) (without the term
$\partial_{i}A_{0}$) and (\ref{d5}). These are the same equations
that can be derived from the canonical Hamiltonian by eliminating
the last term in Eq. (\ref{d3}). However, one should note that the
resulting equation
 \be \ddot{A}_{i}= \nabla^{2}A_{i}-\partial_{i}(\nabla\cdot
 \textbf{A}) \label{d6} \ee
should be considered together with the constraint
 \be \partial_{i}\Pi_{i}= \partial_{i}\dot{A}_{i}=
 \frac{\partial}{\partial t}\left(\nabla\cdot \textbf{A}
 \right)=0. \label{dd6} \ee
In other words, the final answer is any solution of the dynamical
equation (\ref{d6}) with static divergence.

As far as the number of initial constants (in this case initial
fields) is concerned, the dynamical equation (\ref{d6}) brings in
6 initial conditions. However, the constraint (\ref{dd6})
decreases it to 5, in agreement with the previous counting
formula.

It is worth noting that in the Lagrangian formulation the
equations of motion read
 \be L^{\nu}\equiv \partial_{\mu}F^{\mu\nu}=0 \label{d7}. \ee
It is clear that the {\it Eulerian derivatives} $L^{\nu}$ are not
independent functions, since $\partial_{\nu} L^{\nu}=0$.
Therefore, the equations of motion at most can be used to
determine three independent fields $A_{i}$ out of four. However,
in the gauge $A^{0}=0$, $\nu=i$ in Eq. (\ref{d7}) gives the
dynamical equation (\ref{d6}), while for $\nu=0$ the constraint
$\partial/\partial t(\nabla\cdot \textbf{A})=0$ is obtained. One
can check that the consistency of this Lagrangian constraint is
fulfilled identically according to the equations of motion.

We conclude this section by a discussion on the Lorentz gauge.
People are familiar with this gauge in the Lagrangian form
 \be \partial_{\mu}A^{\mu}=\partial_{0}A^{0}+\partial_{i}A^{i}
 =0. \label{d8} \ee
Note that the velocity $\dot{A}^{0}$ can not be obtained in terms
of the phase space variables. On the other hand using Eq.
(\ref{d2}) we have
 \be \dot{A}^{0}=\left\{ A^{0},H_t \right\}=\lambda. \label{d9} \ee
Hence, in the Hamiltonian framework the Lorentz gauge can be
achieved by imposing the GFC
 \be \omega^{(\lambda)}= \lambda +\nabla\cdot\textbf{A}=0
 \label{d10} \ee
which depends on the Lagrange multiplier as well.  This gauge is
in fact equivalent to choosing the Lagrange multiplier in terms of
the physical variables from the very beginning. The dynamics of
the system is then given by the total Hamiltonian
 \be H_t=\int d^{3}x\left[\frac{1}{2}\Pi_{i}\Pi_{i}+
 \frac{1}{4}F_{ij}F_{ij}-A_{0}\partial_{i}\Pi_{i} - \Pi_{0}
 \partial_{i}A_{i} \right], \label{d11} \ee
together with the constraints $\phi_1$ and $\phi_2$. In this way
the canonical equations of motion reproduce Eqs. (\ref{d4}) and
(\ref{d5}) as well as the gauge condition (\ref{d8}) which finally
yield the wave equations for all $A^{\mu}$, as expected. It is
worth noting that the consistency  of $\omega^{(\lambda)}_{2}$,
using Eqs. (\ref{aa1}) and (\ref{d11}) gives
 \be \dot{\omega}^{(\lambda)}
 =\dot{\lambda}+ \nabla^{2}A^{0}, \label{d12} \ee
which from Eq. (\ref{d9}) results in the equation of motion for
$A^0$ (i.e. the wave equation). Therefore the consistency of
$\omega^{(\lambda)}_{2}$ is fulfilled identically.

Now this question may arise: ``which kind of gauge is the Lorentz
gauge, perfect or imperfect?" Remember that in the case of
imperfect gauge fixing, if the primary GFC is conjugate to the
$M$th level constraint, then after $M$ steps of investigating the
consistency of GFC's one would be able to determine the Lagrange
multiplier. Furthermore, $N-M$ remaining constraints should be
imposed on the solutions of the equations of motion. However, in
the case of Lorentz gauge there is no need to follow the
consistency process to find the Lagrange multiplier. On the other
hand, all the constraints are needed to be imposed on the
solutions of the equations of motion, or in other words $M=0$ for
Lorentz gauge. So, roughly speaking, we can say such a gauge is
{\it completely imperfect}. In other words, all of the four fields
$A^{\mu}$ are taken into account within the dynamical equations of
motion (i.e. wave equation) and none of them, or no combination of
them, is omitted according to the gauge.

More accurately, in the case of completely imperfect gauges, the
meaning of GFC's as additional constraints which reduce the
``constraint surface" into the ``reduced phase space", should be
revised. In such systems the gauge orbits disappear by determining
the Lagrange multiplier, rather than by cutting the gauge orbits
by imposing the GFC's. The most interesting fact is that, although
the gauge is fixed, the original Poisson bracket is unchanged. In
other words, no Dirac bracket is needed to describe the algebraic
structure of the physical phase space. Especially, in order to
quantize the theory, all of the eight fields $A^0,A^i,\Pi_0$ and
$\Pi_i$ should be converted to canonical operators, while the
physical subspace of the system is composed of states destroyed by
the first class constraints $\Pi_0$ and $\partial_{i}\Pi_{i}$.
However, this quantized system differs from that obtained by
quantizing the first class system (without gauge fixation) in the
sense that in this case a well defined Hamiltonian, i.e. the
quantized version of (\ref{d11}), is responsible for the evolution
of the system.

We conclude this section by mentioning that since $M=0$ in the
case of completely imperfect gauges, the number of initial
conditions is $2K-N$ for such systems. For electromagnetism in
Lorentz gauge the canonical equations due to the Hamiltonian
(\ref{d11}) bring in 8 initial constants 2 of which are redundant
according to the constraints $\Pi_0$ and $\partial_{i}\Pi_{i}$.

 \section{Polyakov string}

The Polyakov string is introduced \cite{Polchinski,GreenSchWitt}
by the action
 \be S=\frac{1}{4\pi\alpha'}\int d^{2}\sigma
 \sqrt{g}g^{ab}\partial_{a}X^{\mu}\partial_{b}X_{\mu}, \label{e1}
 \ee
where $g_{ab}$ is the world-sheet metric, $g$ is minus the
determinant and $g^{ab}$ is the inverse of $g_{ab}$, $X_{\mu} \ \
\mu=0,1,\cdots ,D-1$ are bosonic fields, and
$\frac{1}{2\pi\alpha'}$ is the tension of the string which can be
taken to be unity. Since $\dot{g}_{ab}
(\equiv\partial_{\tau}g_{ab})$ is absent from the Lagrangian, the
conjugate momentum fields $\pi^{00}$, $\pi^{01} ( =\pi^{10})$, and
$\pi^{11}$ are primary constraints, i.e.
 \be \pi^{ab} \equiv\frac{\partial L}{\partial\dot{g}_{ab}}\approx
 0. \label{e2} \ee
The remaining momenta and the canonical Hamiltonian read
 \ben P_{\mu} &\equiv& \frac{\partial L}{\partial
 \dot{X}^{\mu}}=\sqrt{g}(g^{00}\dot{X}_{\mu}+g^{01}X'_{\mu})
 \label{e3} \\
 H &=& \frac{1}{2}\int d\sigma \left[ \frac{1}{g^{00}} \left(
 \frac{P_{\mu}}{\sqrt{g}}-g^{01}X'_{\mu} \right)^2 -g^{11}X'^2
 \right]\nonumber \\  &=& \frac{1}{2}\int d\sigma \left[
 \frac{1}{\sqrt{g}g^{00}} \left( P^2+X'^2 \right) -
 \frac{2g^{01}}{g^{00}} P.X' \right], \label{e4} \een
where ``dot" and ``prime" represent differentiating with respect
to $\tau$ and $\sigma$ respectively. Then one should investigate
the consistency of the primary constraints $\pi^{ab}$. Since the
canonical Hamiltonian depends on the string variables ($X$ and
$P$) only through the functions $(P^2+X'^2)$ and $P.X'$, the
consistency procedure will give some functions of the metric
variables ($g_{ab}$ or $g^{ab}$) times the above functions, which,
as we will see in the following, have weakly vanishing Poisson
brackets with each other and with the Hamiltonian. Therefore,
without going through detailed calculations, one can guess that
there exists no further constraint after the second level.

These observations suggest a change of variables from the original
metric components to some suitable combinations of them. Eq.
(\ref{e4}) shows that the following variables are adequate,
 \ben N_1 &=& \frac{1}{\sqrt{g}g^{00}} \label{e5} \\ N_2 &=&
 -\frac{g^{01}}{g^{00}}\label{e6}\\ N_3 &=&\sqrt{g}\label{e7}.\een
The variable $N_3$ is also dictated by the fact that the action is
independent of the scale of the metric which can be given by
$\sqrt{g}$. The metric can be written in terms of $N_i$ as
 \be g_{ab}= \left( \bea{cc} N_3N_1[1-(N_2/N_1)^2] & -N_3N_2/N_1
 \\ -N_3N_2/N_1 & -N_3/N_1 \eea \right). \label{e13} \ee
Writing the action in terms of $N_i$, it is clear that their
conjugate momenta $\Pi^i$ are primary constraints. Then using the
total Hamiltonian
 \be H_t=H +\int d\sigma \sum_{i=1}^{3}\lambda_{i}\Pi^{i},
 \label{ee1} \ee
 where
 \be H=\frac{1}{2}\int d\sigma \left[ N_1 (P^2+X'^2)+N_2(2
 P.X')
 \right], \label{e8} \ee
the consistency of $\Pi^3$ is satisfied trivially, while the
consistency of $\Pi^1$ and $\Pi^2$ give respectively the following
secondary constraints
 \be \bea{l} \Phi_{1}=\frac{1}{2}(P^2+X'^2) \\ \Phi_2=P.X'.
 \eea \label{ee2} \ee
The Poisson brackets of the secondary constraints read
 \ben \{\Phi_1(\sigma ,\tau),\Phi_1(\sigma' ,\tau)\}&=&\partial\delta
 (\sigma-\sigma') \left[P(\sigma',\tau).X'(\sigma,\tau)+P(\sigma,
 \tau).X'(\sigma',\tau)\right] \label{e10} \\ \{\Phi_1(\sigma ,\tau),
 \Phi_2 (\sigma' ,\tau)\}&=& \partial\delta
 (\sigma-\sigma') \left[ X'(\sigma',\tau).X'(\sigma,\tau)+P(\sigma,
 \tau).P(\sigma',\tau)\right] \label{e11}\\\{\Phi_2(\sigma ,\tau),
 \Phi_2(\sigma' , \tau)\}&=&  \partial\delta
 (\sigma-\sigma') \left[P(\sigma',\tau).X'(\sigma,\tau)+P(\sigma,
 \tau).X'(\sigma',\tau)\right], \label{e12}   \een
which vanish on the constraint surface. Therefore no other
constraint emerges. Using the language of the chain by chain
method, we have derived the following three first class constraint
chains
 \be \bea{lll} \Pi^1 \hspace{5mm} &\Pi^2 \hspace{5mm}&\Pi^3 \\
 \Phi_1 \hspace{5mm}&\Phi_2 \hspace{5mm}& \\ \eea .\label{e9}\ee

The third chain in (\ref{e9}) contains only one element, $\pi^3$,
i.e. the generator of gauge variation of $N_3$ which only changes
the scale of the world-sheet metric. This is the well-known Weyl
symmetry of the Polyakov string. The remaining constraints in
(\ref{e9}), i.e. $(\Pi^1,\Pi^2;\Phi_1,\Phi_2)$, generate the
reparametrizations in a more complicated way, which is not of our
interest here.

Next we want to investigate the problem of gauge fixing in the
Polyakov string. As we observed, all three independent components
of the world-sheet metric are gauge variables and can be
determined by fixing the gauge. For example one may assume the
world-sheet metric to be flat, i.e. $g_{ab}=\eta_{ab}$,
corresponding to $N_1=-1$, $N_2=0$ and $N_3=1$. This choice of
gauge kills the reparametrization, as well as the Weyl invariance
of the Polyakov action. However, since the above CFC's are
conjugate to the first level constraints $\Pi^i$, the
corresponding gauge is an imperfect one. Therefore, the second
level constraints $\Phi^1$ and $\Phi^2$ remain unfixed and should
be imposed after all on the solutions of the classical equations
on motion. Also at the quantum level the Virosoro constraints
$\Phi^1$ and $\Phi^2$ as first class constraints should kill the
physical states. Quantization of the bosonic string in this gauge
is known to the string theorists as ``the old covariant
quantization" \cite{GreenSchWitt}.

Suppose one had considered, from the very beginning, the
world-sheet metric is flat. Then one would have a different theory
without any first class constraint. In such a theory one
encounters a traceless energy- momentum tensor, but there will be
no justification for imposing the Virosoro constraints $\Phi^1$
and $\Phi^2$ which imply vanishing of the energy- momentum tensor.
It is essential to distinguish between the gauge fixed theory of
the bosonic string coupled to two dimensional gravity (i.e.
Polyakov string) and the theory of bosonic string living on a flat
world-sheet.

Although the general covariance of the world-sheet disappears in
the above imperfect gauge, this gauge has the advantage of keeping
the target space Lorentz covariance of the fields. This is in
contrast with the famous light cone gauge, in which the Lorentz
covariance of the coordinate fields $X^{\mu}$, as well as the
general covariance of the world-sheet, are destroyed. The light
cone gauge is in some sense a perfect gauge. The important point
concerning a perfect gauge is that it should first fix the gauge
freedom by determining the undetermined Lagrange multipliers; and
next, the consistency conditions of the corresponding GFC's should
give as many GFC's as the constraints, so that there remain no
more unfixed first class constraints which may act as the
generator of any residual symmetry.

The light cone gauge may be introduced, in terms of the variables
$N_1$, $N_2$, $N_3$, $X^{\pm}\equiv (X^0\pm X^1)/\sqrt{2}$ and
$X^i$, $i=2\cdots D$, by the following GFC's
 \ben \Omega_{1}&=& N_1+1 \label{e14}\\ \Omega_2&=& N'_2 \label{e15}
 \\ \Omega_3&=& N_3-1 \label{e16}
 \\ \omega_2 &=& X^+ -a\tau -b \label{e17}\een
These are four GFC's, while we have five FCC's. Therefore one more
GFC is needed to fix the gauge perfectly. Consistency of
$\Omega_3$ in (\ref{e16}) determines $\lambda_3$ in (\ref{ee1}) to
be zero and fixes the Weyl gauge transformation generated by
$\Pi^3$. To fix the reparametrizations, let us first consider the
consistency of $\omega_{2}$ in Eq. (\ref{e17}). Using Eqs.
(\ref{aa1}) and (\ref{e14}-\ref{e17}) we have
 \be \dot{\omega}_{2}\equiv \omega_{1}=P_{-}-a
 \label{e18} \ee
This GFC completes the set of required GFC's to fix all the gauge
transformations. The consistency of $\omega_{1}$ gives
$(N_1X'_-)'+(N_2P_-)'$, which vanishes identically by imposing
(\ref{e14}-\ref{e18}) and gives no further result. Finally the
consistency of $\Omega_{1}$  and $\Omega_2$ determines the
remaining Lagrange multipliers  as $\lambda_1=\lambda'_2=0$.
Imposing the boundary conditions determines the constant value of
$\lambda_2$ to be zero.

The coordinate $X^+$ as well as the momentum $P_-$ are determined
according to GFC's $\omega_{2}$ and $\omega_{1}$. Note that, using
the constraints $\Phi_{1}$ and $\Phi_{2}$, the conjugate fields
$P_+$ and $X^-$ may be determined in terms of the transverse
coordinates and momenta (i.e. $X^i$ and $P_i$) as follows
 \ben  P_+ &\approx& \frac{1}{2a}(P^iP_i+{X^i}'{X_i}'), \label{e19}\\
 {X^-}' &\approx& \frac{1}{a}(P_i{X^i}').  \label{e20} \een
In this way all the gauges are fixed and there remain only
transverse coordinates as  the physical fields which possess
independent dynamics in the classical level. This result may be
compared with the light cone gauge in the case of relativistic
point particle in which $X^-$ is remained as a dynamical
coordinate (see Eq. \ref{c26}). To quantize the theory one should
find the Dirac brackets due to these 10 constraints (i.e. 5 FCC's
and 5 GFC's given above). It is not difficult to see that
 \be \{X^i(\sigma,\tau),P_j(\sigma',\tau)\}_{DB}=\delta^{i}_{j}
 \delta(\sigma-\sigma'), \label{e21} \ee
while all other Dirac brackets vanish. Therefore the system may be
easily quantized (after imposing suitable boundary conditions) by
quantizing just the transverse coordinates with no need to impose
subsidiary conditions on the physical states. The details may be
found in any text book on string theory.

\section{concluding remarks}

We showed, in this paper, that the chain by chain method in
constructing the constraint structure of a gauge theory provides a
suitable framework for classifying different types of gauge
fixing. From this point of view we introduced perfect and
imperfect gauge fixings.

Perfect gauge fixing happens when the gauge fixing conditions are
chosen to be conjugate to the last elements of the first class
chains. In this category of gauges, the consistency of primary
gauge fixing conditions produces newer ones. Repeating this
procedure leads to the emergence of an adequate number of gauge
fixing conditions such that every gauge generator (i.e. first
class constraint) has its conjugate among the set of assumed and
produced gauge fixing conditions. Therefore, the gauge would be
fixed perfectly, so that no residual symmetry would be generated
by the unfixed gauge generators. The relativistic point particle
in temporal and light cone gauges, electromagnetism in Coulomb
gauge, and the Polyakov string in light cone gauge are shown to be
examples of perfect gauges.

Imperfect gauge fixings concern cases in which the primary gauge
fixing conditions are proposed to be conjugate to some first class
intermediate constraints in the constraint chains which we call
the gauge fixing level. The consistency of primary gauge fixing
conditions produce newer ones which are conjugate to the
constraints preceding the gauge fixing level. Therefore, the
constraints succeeding this level remain unfixed and may still
generate residual symmetries. Hence, it is necessary to take into
account the remaining unfixed constraints as subsidiary conditions
which should be imposed, at the classical level, on the solutions
of the equations of motion; and should kill, at the quantum level,
the physical states. In other words, imposing an imperfect gauge
on the original Hamiltonian or Lagrangian is not enough; it is
also necessary to follow up the history of the constraint
structure of the system and impose the original constraints on the
solutions of the gauge fixed system. Relativistic point particle
in the gauge which determines the einbin variable,
electromagnetism in the vanishing potential gauge ($ A^0=0$), and
the polyakov string in the old covariant gauge are examples of
imperfect gauges.

An interesting observation in studying electromagnetism is that
the Lorentz gauge has a special character which we call a
completely imperfect gauge. In this system, one fixes the gauge by
determining the Lagrange multipliers from the very beginning.
Therefore, although the gauge freedom is fixed directly, all the
first class constraints are remained unfixed and should be
considered as subsidiary conditions.

We also had a discussion on the number of initial constants in
different gauges. We showed that this number is the smallest in
the case of a perfect gauge.

\textbf{Acknowledgment}

The author thanks H. Soltan-Panahi for his valuable discussions
and useful ideas and A. E. Mosaffa as well as F. Loran for reading
the manuscript.

\end{document}